\documentclass[aps,prl,twocolumn,longbibliography]{revtex4-1}

\usepackage{amssymb}
\usepackage{amsbsy}
\usepackage{amsmath}
\usepackage{graphicx}
\usepackage{graphics}
\usepackage{setspace}
\usepackage{array}
\usepackage{color}
\usepackage{fontenc}
\usepackage{textcomp}
\usepackage{bm}
\usepackage{float}
\usepackage[bookmarks=false,linkcolor=blue,urlcolor=blue,colorlinks,citecolor=blue]{hyperref}

\newcommand{\vex}[1]{\bm{\mathrm{#1}}}

\newcommand{\bsub}{\begin{subequations}}
\newcommand{\esub}{\end{subequations}}

\graphicspath{{../Figures/}}

\begin{document}
\title{Vortex and Surface Phase Transitions in Superconducting Higher-order Topological Insulators}
\author{Sayed Ali Akbar Ghorashi$^1$, Taylor L. Hughes$^2$, Enrico Rossi$^1$}
\affiliation{$^1$Department of Physics, William $\&$ Mary, Williamsburg, Virginia 23187, USA}
\affiliation{$^2$Department of Physics and Institute for Condensed Matter Theory,
University of Illinois at Urbana-Champaign, IL 61801, USA}

\date{\today}

\newcommand{\be}{\begin{equation}}
\newcommand{\ee}{\end{equation}}
\newcommand{\bea}{\begin{eqnarray}}
\newcommand{\eea}{\end{eqnarray}}
\newcommand{\h}{\hspace{0.30 cm}}
\newcommand{\vs}{\vspace{0.30 cm}}
\newcommand{\n}{\nonumber}
%----------------------------------------------------------------------------------
%%%%%%%%%%%%%%%%%%%%%%%%%%%%%%%%%%%%%%%%%%%%%%%%%%%%%%%%%%%%%%%
\begin{abstract}
Topological insulators (TIs) having intrinsic or proximity-coupled s-wave superconductivity host Majorana zero modes (MZMs) at the ends of vortex lines. The MZMs survive up to a critical doping of the TI at which there is a vortex phase transition that eliminates the MZMs. In this work, we show that the phenomenology in higher-order topological insulators (HOTIs) can be qualitatively distinct. In particular, we find two distinct features. (i) We find that vortices placed on the gapped (side) surfaces of the HOTI, exhibit a pair of phase transitions as a function of doping. The first transition is a surface phase transition after which MZMs appear. The second transition is the well-known vortex phase transition. We find that the surface transition appears because of the competition between the superconducting gap and the local $\mathcal{T}$-breaking gap on the surface. (ii) We present numerical evidence that shows strong variation of the critical doping for the vortex phase transition as the center of the vortex is moved toward or away from the hinges of the sample. We believe our work provides new phenomenology that can help identify HOTIs, as well as illustrating a promising platform for the realization of MZMs.
\end{abstract}
\maketitle

\emph{Introduction}.--In the past decade there has been an explosion of interest in new forms of topological matter, driven by the discoveries of topological insulators and gapless topological semimetals\cite{hasan2010,qi2011,Chiu2016}. The search for Majorana zero modes (MZMs) has been at the heart of it, due to its promising applications in developing the building blocks of topological quantum computation\cite{Nayakrev}. In a seminal work, Fu and Kane\cite{FuPRL2008} showed that a MZM can be trapped in the core of a vortex when an s-wave superconductor proximitizes a TI with gapless, Dirac surface states. Later, Hosur et. al. \cite{hosur2011} demonstrated that these MZMs can actually survive up to some critical doping of the TI bulk bands beyond the band edges\cite{hughesVPT11,hughesVPT13,expRevVortex}. \\
\indent Recently, some aspects of topological phases of matter have received newborn attention after the introduction of so-called higher-order topological phases \cite{Benalcazar2017-1,Benalcazar2017-2,Schindler2018-2,Schindler2018-1,Langbehn2017,Song2017}, which has spawned numerous works in the last few years \cite{GhorashiHOTSC2019,EzawaPRL2018,EzawaPRB2018,Khalaf2051,CAlugAru2018,Kheirkhah2019,ZhongWang2018,Loss2018,wanghughes18,FanZhang2018,Nori2018,ZhuPRL2019,DasSarmaPRL2019,Fulga2019,Yan2019,Hsu2019,Zhang2019}.
%
%Unlike  conventional topological insulators,
An $n^{th}$-order topological phase of a system of dimension $d$
possesses gapless boundary modes on the $d-n$ dimensional boundaries
with $1<n\leq d$, unlike a conventional topological phase for which $n=1$.
%
%where $d$ is the system's dimension.
Since its first theoretical discovery\cite{Benalcazar2017-1}, there have been experimental realizations of higher-order topological insulators (HOTIs) in several  meta-material contexts \cite{Peterson2018,Noh2018,Serra-Garcia2018,Imhof2018}, as well as evidence in solid state electronic materials\cite{Schindler2018-2}\\
\begin{figure}[htb]
    \centering
    \includegraphics[width=0.51\textwidth]{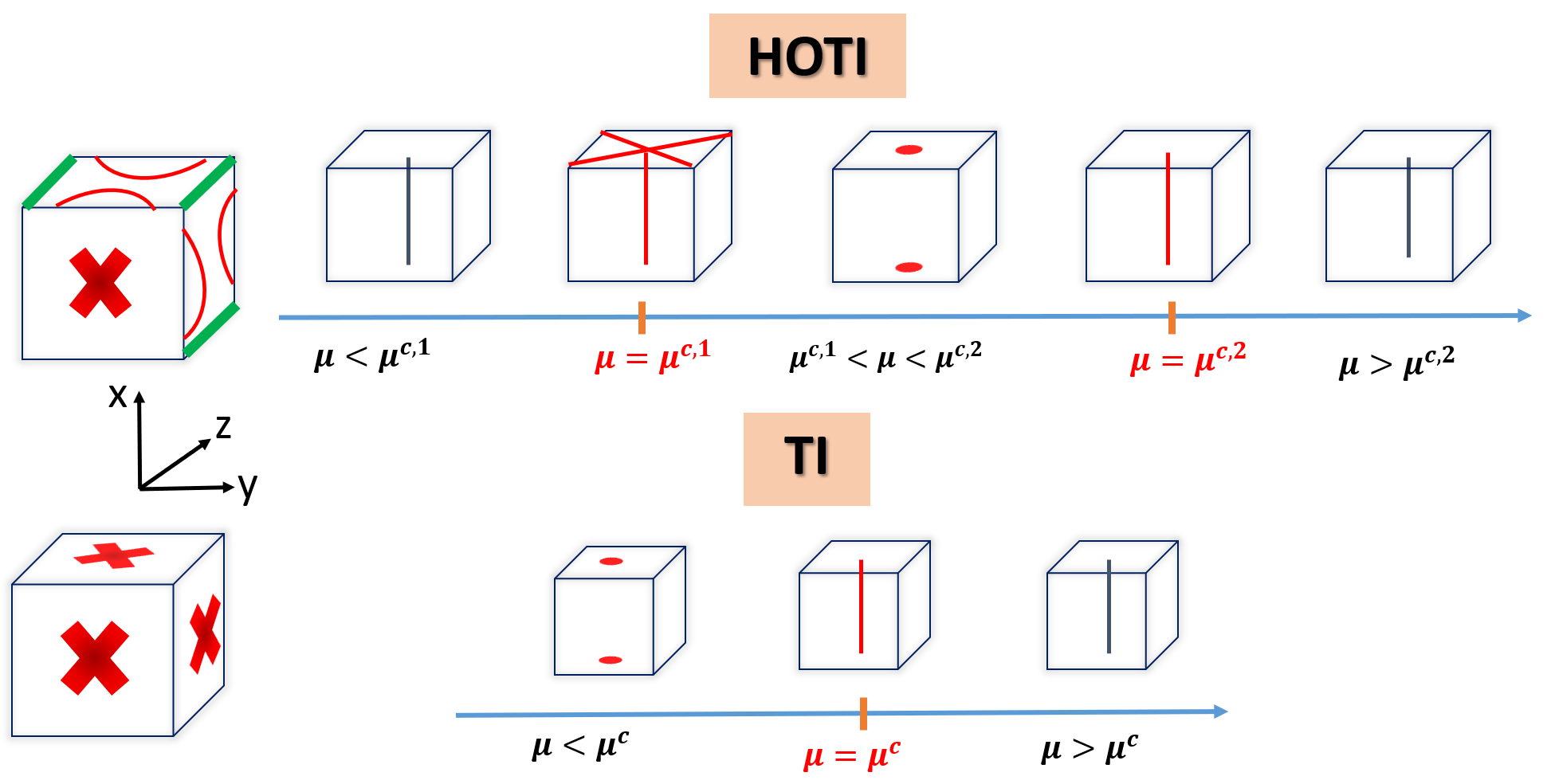}
    \caption{A schematic picture depicting the evolution of the vortex bound states with doping. The low-energy 1D vortex spectrum can be utilized to distinguish between HOTI (with chiral hinge states) from a conventional TI. By tuning of doping at the critical doping $\mu^{c,1}$, the gapped surfaces of HOTIs undergo a surface phase transition and stable MZMs only appear for $\mu^{c,1}<\mu<\mu^{c,2}$.}
    \label{fig:adpic}
\end{figure}
\indent In light of this previous work we can ask a natural question: can proximitized (or intrinsically superconducting) 3D HOTIs exhibit MZMs in their vortices? In this article we answer this question in the affirmative, but reveal the appearance a surface phase transition tuned by changes in the chemical potential that distinguishes the HOTI from the ordinary TI phenomenology. We identify the general condition for the appearance of the surface transition making our results relevant to
a large class of higher-order 3D topological systems. Recent works \cite{exhoti1,exhoti2,exhoti3,exhoti4,exhoti5,exhoti6,exhoti7,wang2020evidence,gray2019evidence,DasSarmaPRL2019,hotscironDasSarmaPRL2019} have pointed out that some of the magnetic and axion insulators as well as iron-based superconductors might be related to higher-order topological phases. Also, many experiments \cite{vortexexp1,vortexexp2,vortexexp3,vortexIron3,Iron1} have studied the physics of vortices in iron-based superconductors. Therefore, these developments suggest that our results might apply to systems that are already actively studied experimentally.

We begin with a conventional 3D $\mathcal{T}$-invariant topological insulator. From here, Ref. \onlinecite{Schindler2018-1} predicts that one can form a 3D HOTI having chiral hinge states if one adds a $C^z_4\mathcal{T}$ symmetric term. This term acts to gap out the surfaces perpendicular to the $x$ and $y$ directions, but leaves the $z$-surfaces gapless. When proximitized by s-wave superconductivity, the $z$-surfaces behave as in an ordinary TI, as long as any vortices are far from any (gapless) hinges of the sample. However, vortices on the side surfaces show a different phenomenology. Besides the large critical doping $\mu^{c,2}$ that marks the known vortex phase transition (VPT)\cite{hosur2011}, we find a new lower critical doping $\mu^{c,1}$ (Fig.~\ref{fig:adpic}). For chemical potentials smaller than $\mu^{c,1}$ no stable MZMs are bound to vortices on the side surfaces. The lower critical point represents a topological surface phase transition resulted from the competition between superconducting gap and the $C_4\mathcal{T}$ symmetric term. Therefore, we find that stable, vortex-bound MZMs do exist on the side surfaces for the range $\mu^{c,1}<\mu<\mu^{c,2}$ (Fig.~\ref{fig:adpic}).

 We illustrate these transitions, in comparison with an ordinary proximitized TI, in Fig. \ref{fig:adpic}. We believe that our results clearly distinguish the vortex phenomenology in proximity induced TIs and HOTIs, and can be employed as a promising approach for identifying HOTIs as well as platform for the realization of MZMs.

\emph{Model}.-- We begin with a lattice model for a 3D superconducting chiral HOTI with s-wave singlet pairing. Let $\Psi^T_{\vex{k}}=\left(\psi_{\vex{k}}, \psi^{\dagger}_{-\vex{k}}\right)$ where $\psi^T_{\vex{k}}=\left(c_{\uparrow},\,c_{\downarrow}\,d_{\uparrow},\,d_{\downarrow}\right)$ and $(c,d)$ represent different orbitals while $\uparrow/\downarrow$ are spin.  The Bogoliubov-de-Gennes (BdG) Hamiltonian for our model is is $H_{SC}=\sum_{\vex{k}}\Psi^{\dagger}_{\vex{k}}h^{BdG}_{\vex{k}}\Psi_{\vex{k}}$ where
\begin{eqnarray}\label{hambdg}
h^{BdG}_{\vex{k}}=\begin{bmatrix}
   H_{HOTI}(\vex{k})-\mu & \Delta \\
   \Delta^* & -H^T_{HOTI}(-\vex{k})+\mu\\
\end{bmatrix}
\end{eqnarray}
and \begin{align}
    &H_{HOTI}(\vex{k})=\,\left(M+t_0\sum_i \cos (k_i)\right)\kappa_z\sigma_0\cr
+&\,t_1\sum_i\sin (k_i)\kappa_x\sigma_i\,+\,t_2\large(\cos (k_x) - \cos (k_y)\large)\kappa_y\sigma_0.
\label{eqhoti}
\end{align}
$\kappa$, and $\sigma$ are Pauli matrices in orbital and spin spaces, respectively and we set lattice constant to one.
For $t_2=0$, Eq.~(\ref{eqhoti}), represents the well-known model of 3D time-reversal invariant TI, for $1<|M|<3$ (strong TI) and $0<|M|<1$ (weak TI). For $t_2\neq 0$ the system can be in a  3D chiral HOTI phase when the TI is tuned to a strong TI phase\cite{Schindler2018-1}. The term proportional to $t_2$ breaks time-reversal and $C_4$ rotational symmetry while preserving their  product: $(C_4\mathcal{T})H^*_{HOTI}(k_x,k_y,k_z)(C_4\mathcal{T})^{-1}=H_{HOTI}(k_y,-k_x,-k_z)$, $C_4=\kappa_0e^{-i\pi\sigma_z/4}$ and $\mathcal{T}=\kappa_0\sigma_y$.
%
% CHANGE
%
As a result, $x,y$-surfaces are gapped while the $z$-surface remains gapless.
The model of Eq.~\eqref{eqhoti} is ideal to understand the differences between the vortex phase transition in a
superconducting TI and a superconducting HOTI given that:
(i)  it is a prototypical model of chiral HOTI that reduces to a standard TI for $t_2\to 0$;
(ii) it has two gapless surfaces and therefore allows us to understand how the vortex' spectrum
     is affected by the nature, gapless or not, of the surfaces of the HOTI, and
     to understand the differences, for the vortex spectrum, between the gapless surfaces of the TI
     and the gapless surfaces of the  HOTI.
We also considered other {\em standard} HOTI models \cite{ChiralHOTISM,ChiralHOTIold1,ChiralHOTIold2,Langbehn2017,Khalaf2051,Ezawamagnetic2018}, see SM \cite{sm}, and found qualitative similar results.
For the pairing term (assumed to be s-wave) we choose $\Delta=\delta_0\kappa_0 \sigma_y$ where $\delta_0$ is the strength of the pairing.
We choose this pairing for two reasons:
(i)  among all the possible s-wave pairing, \cite{sm,Sato,Fu2010,Roy-Ghorashi2019}, it is the only one for which the bulk remains gapped as the chemical potential is varied,
(ii) it is the pairing that was considered in the study of the vortex phase transition in a superconducting TI~\cite{hosur2011}.

In the presence of superconductivity, the HOTI will develop chiral Majorana hinge modes, with a pair of chiral modes on hinges parallel to $z$ that split when they intersect the top and bottom surfaces\cite{queiroz2019}.
\begin{figure}[h]
\includegraphics[width=0.26\textwidth]{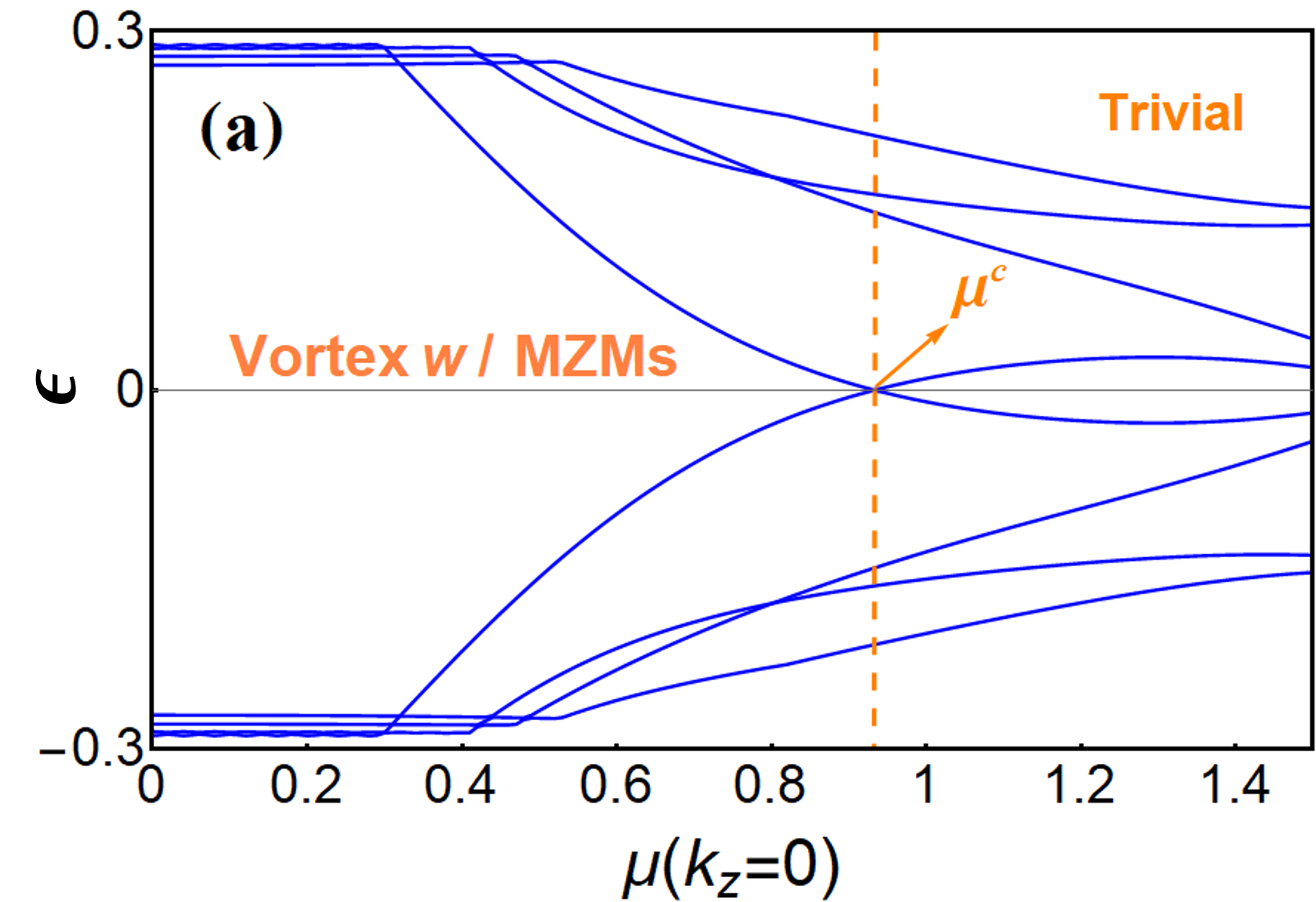}\,\includegraphics[width=0.26\textwidth]{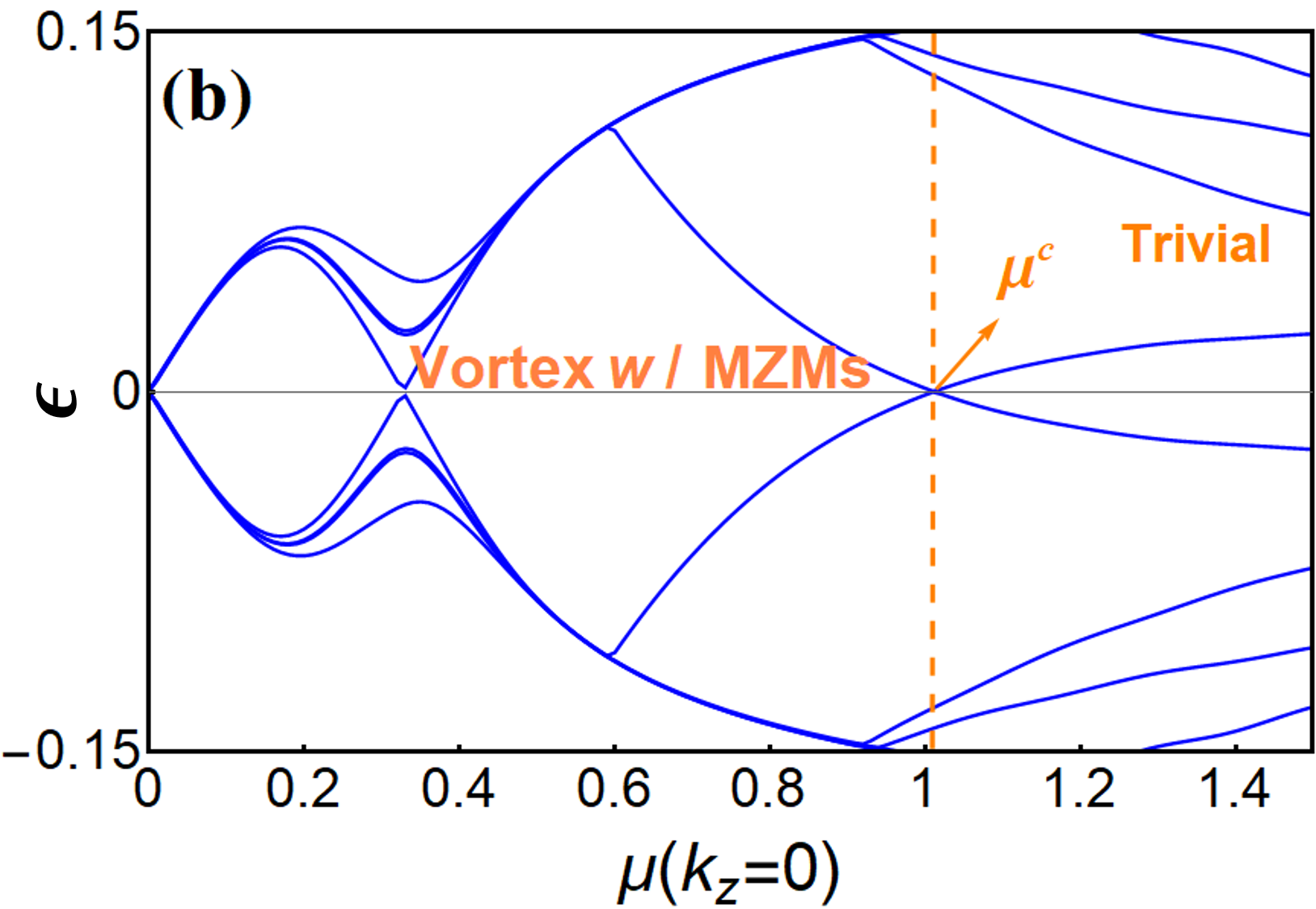}
\includegraphics[width=0.26\textwidth]{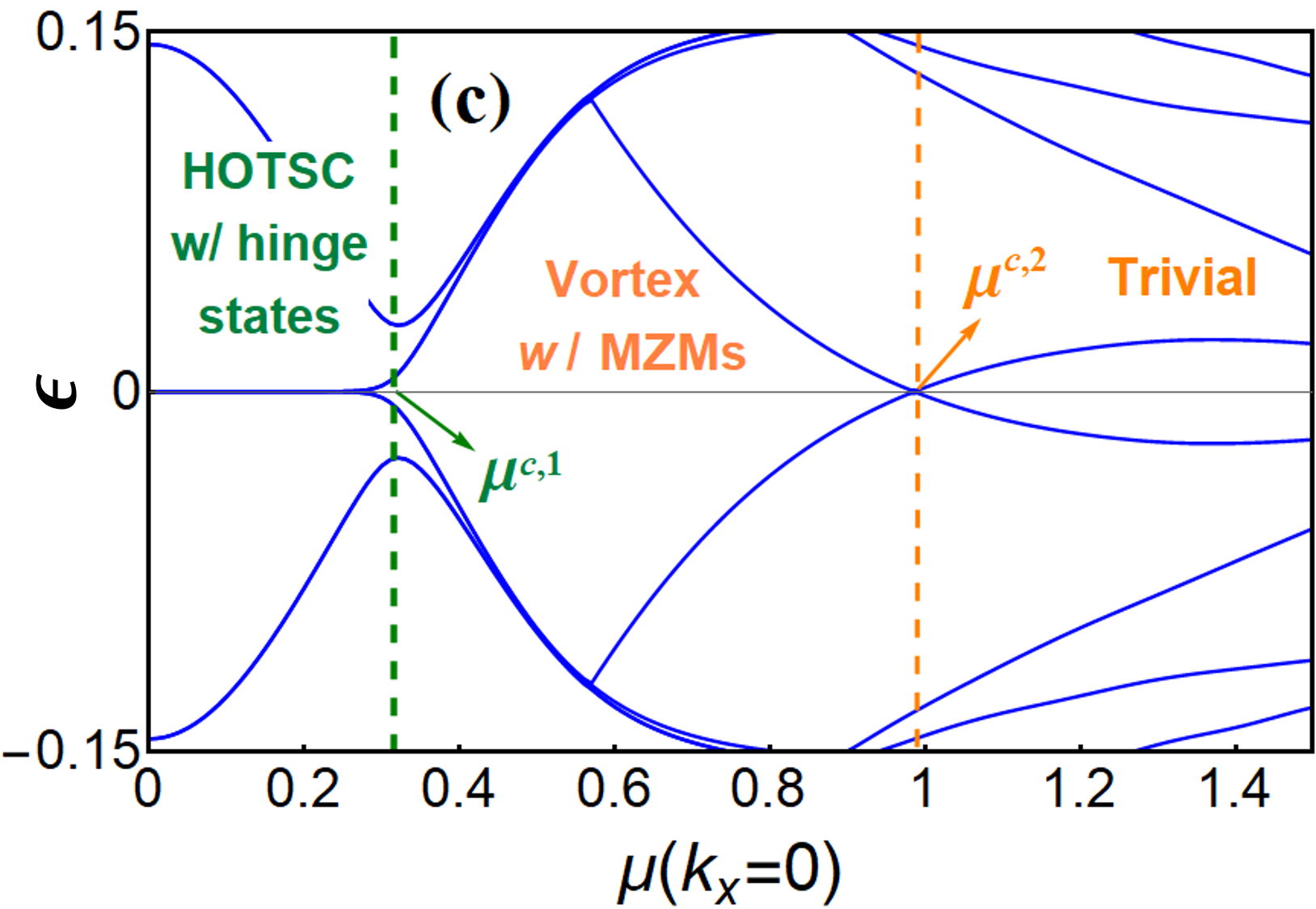}
     \caption{VPT spectrum versus $\mu$ for (a) a TI with vortex line parallel to $z$ (b) a HOTI with a vortex line parallel to $z$ (c) a HOTI with a vortex line parallel to $x$. For (a,b) the system size is $40 \times 40$ in the plane perpendicular to the vortex line. In (c) the system size is $100 \times 100.$   Model parameters $M=-2.5, t_0=1, t_1=1, t_2=1,\delta_0=0.3$ are used.}
     \label{fig:VPTvsk}
 \end{figure}
We can implement a vortex line in a plane using a spatially dependent pairing term $\Delta(\vex{r})=\Delta\tanh(r/\xi_0)e^{i\phi_0}$, where $r=\sqrt{x_i^2+x_j^2}$, $\phi_0=\arctan(x_j/x_i),$ and we choose $\xi_0=1$.
We ignore any contributions from the vector potential and Zeeman term from the field used to induce the vortex as in Ref. \onlinecite{hosur2011}.
The model for the strong TI has cubic symmetry so inserting a vortex on surfaces normal to $x,y,$ or  $z$ gives rise to the same physics. However, this is not the case for HOTIs, as cubic symmetry is broken to an axial symmetry that distinguishes $z$ from $x$ and $y$. From the $C_4\mathcal{T}$ symmetry one expects similar behavior for vortices along $x$ and $y$ (up to a time-reversal transformation), but the $z$-direction can be distinct.\\

\begin{figure*}[t]
    \includegraphics[width=0.25\textwidth]{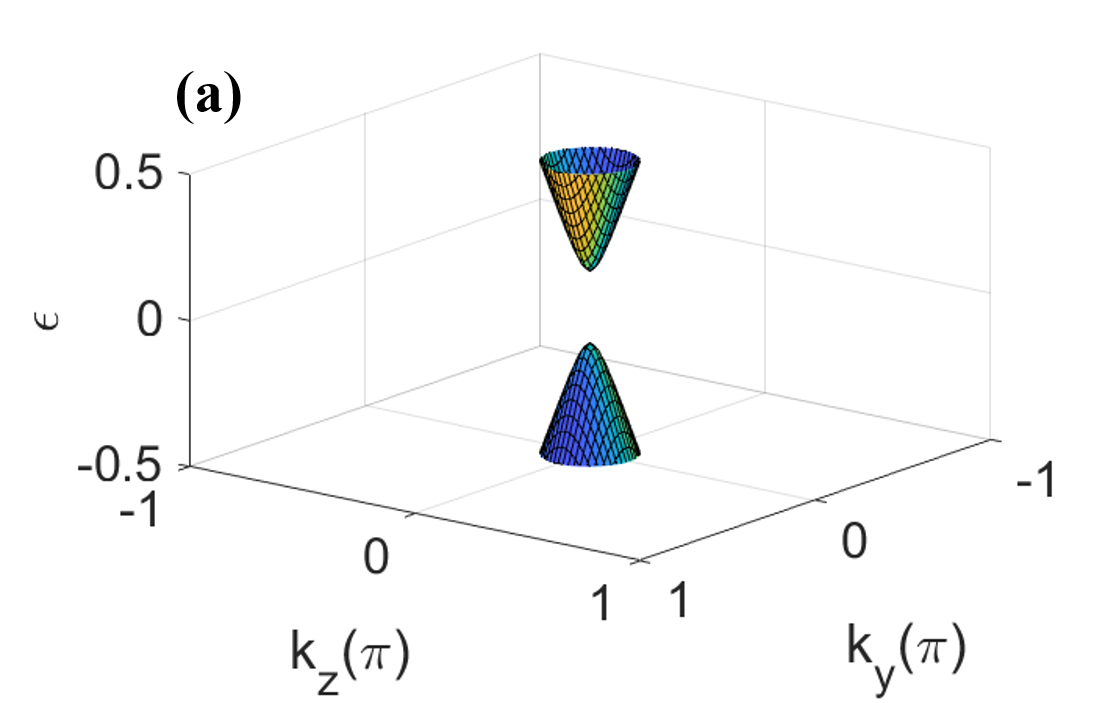}\includegraphics[width=0.25\textwidth]{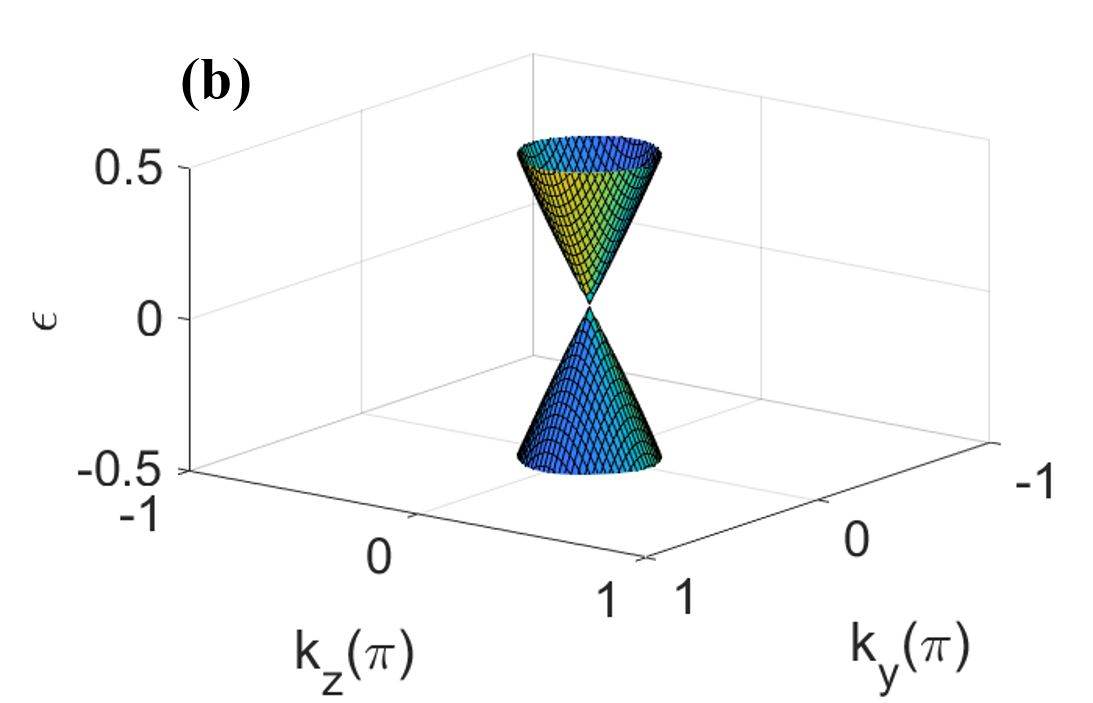}\includegraphics[width=0.25\textwidth]{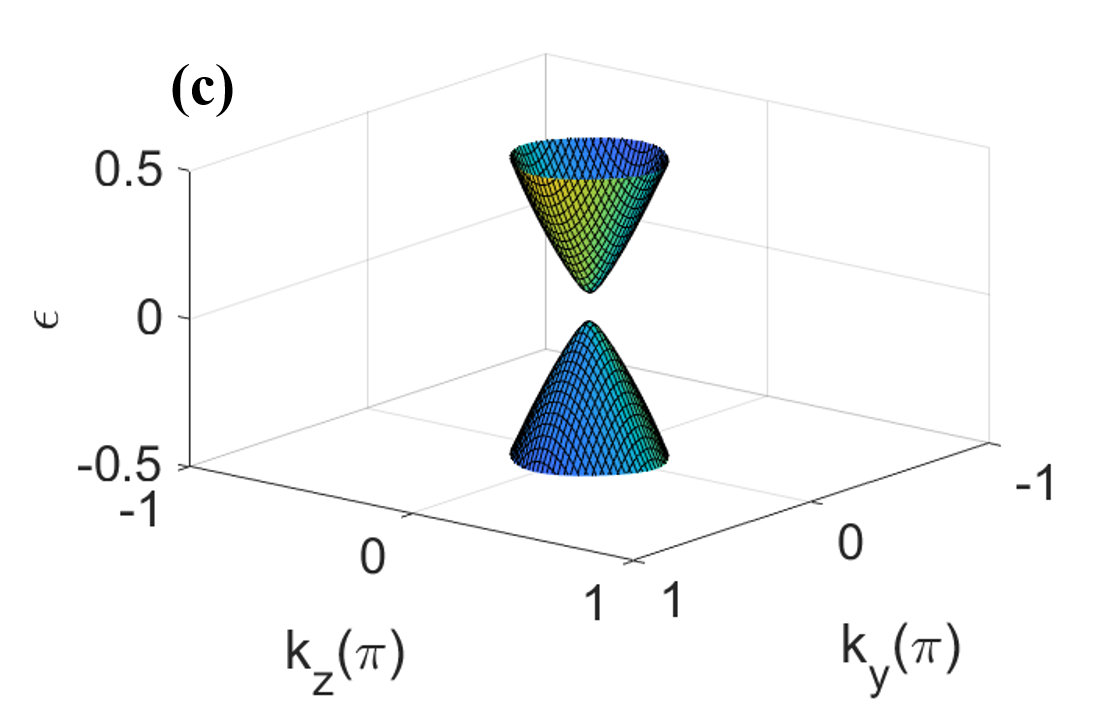}\includegraphics[width=0.26\textwidth]{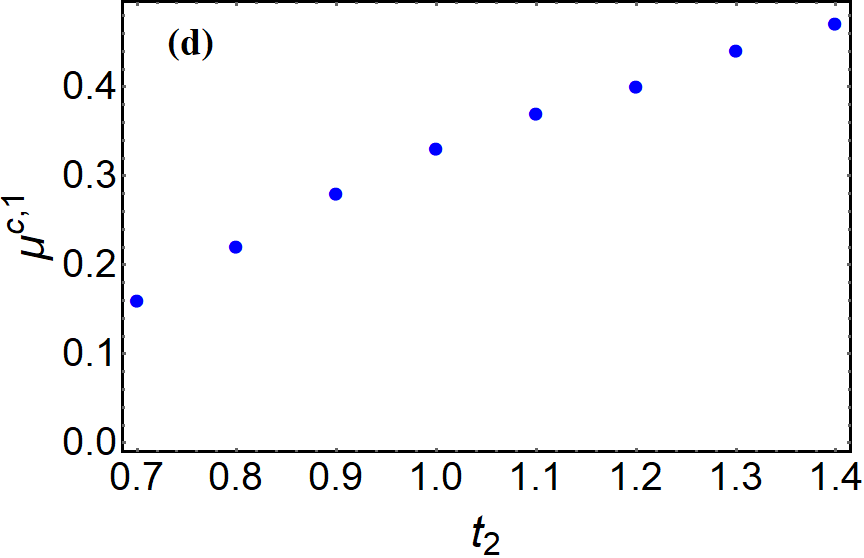}\\
    \includegraphics[width=0.24\textwidth]{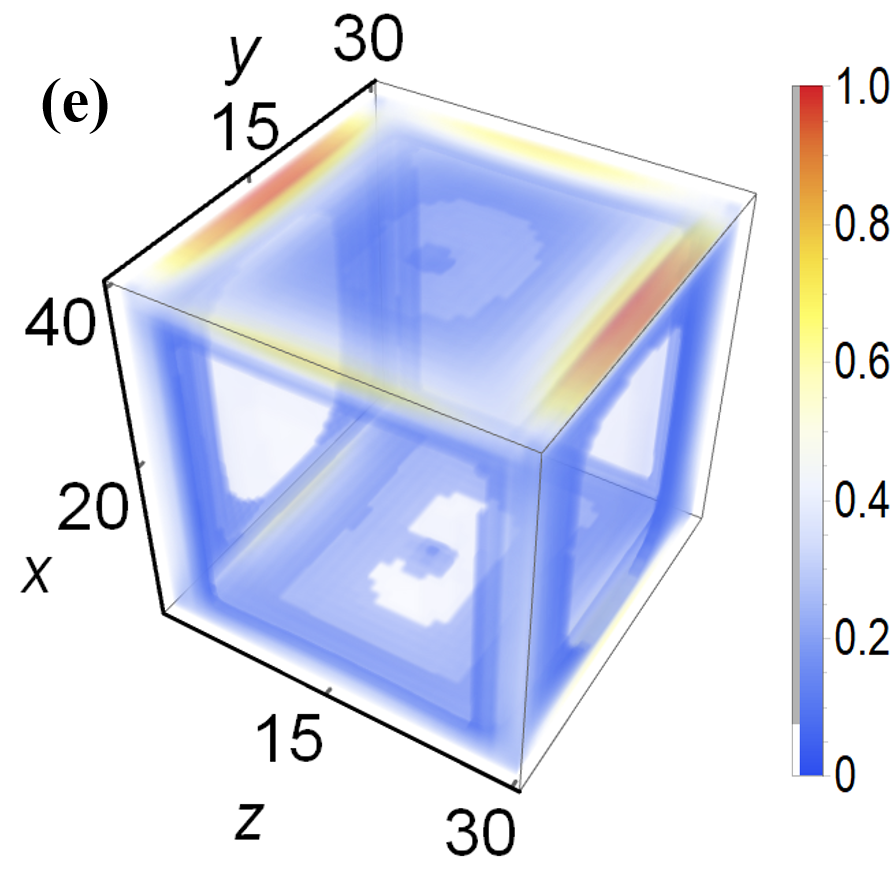}\,\,\,\includegraphics[width=0.24\textwidth]{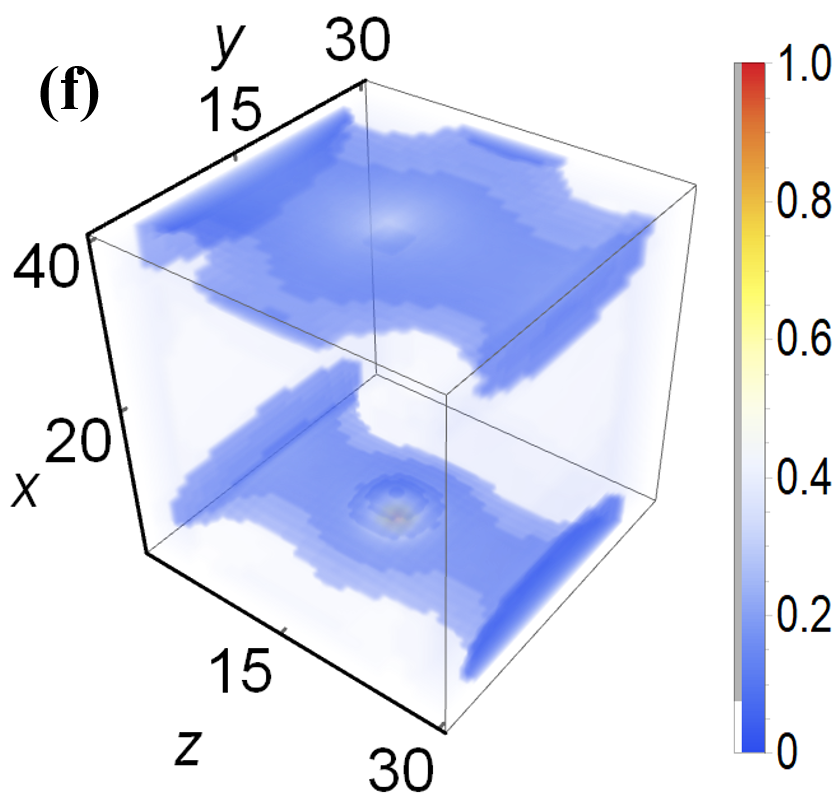}\,\,\,\,\includegraphics[width=0.24\textwidth]{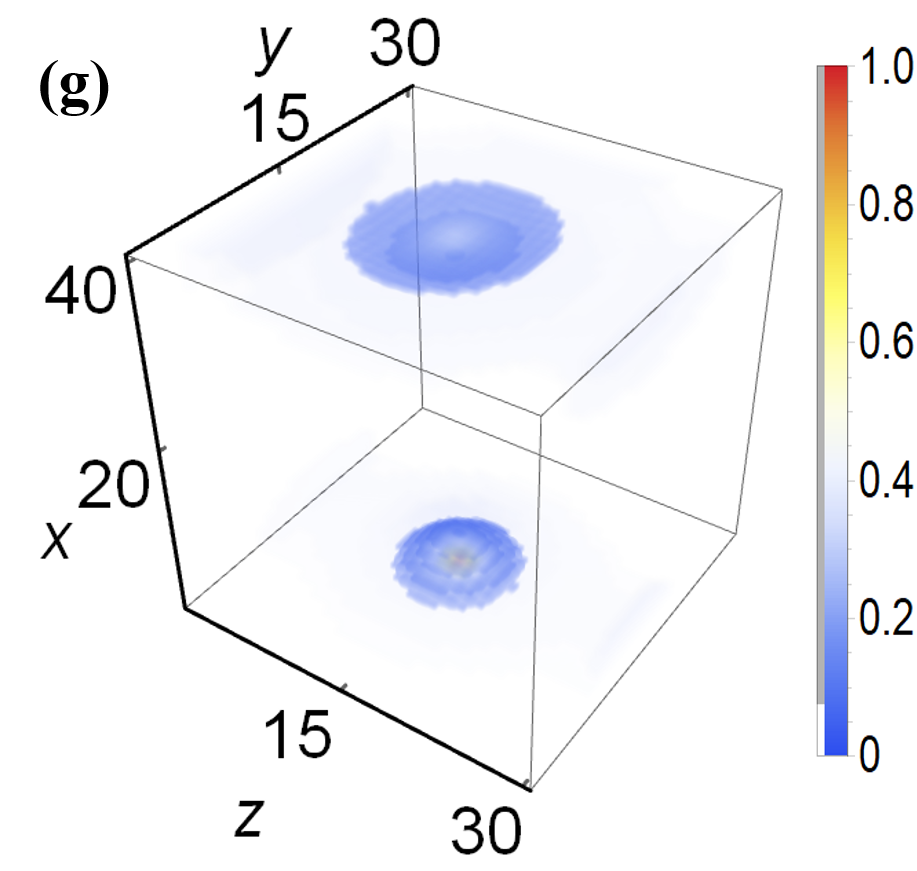}\,\,\,\,\includegraphics[width=0.24\textwidth]{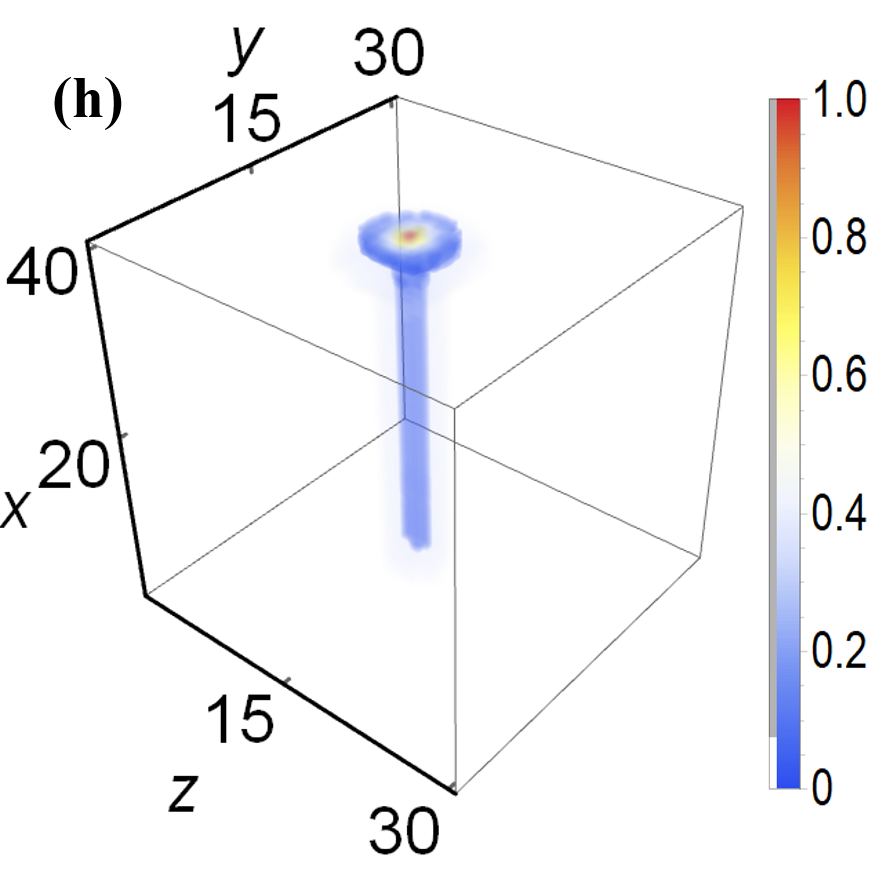}
    \caption{Surface phase transition in a superconducting HOTI. (a-c) The evolution of surface states in the $y-z$ plane for $\mu=0.1,0.33,0.4$, respectively. A surface phase transition occurs at $\mu^{c,1}\sim 0.33$. (d) $\mu^{c,1}$ vs $t_2$. (e-h) The evolution of the probability density profile of the lowest energy states when a vortex is inserted perpendicular to the $y-z$ surface and $\mu$ is tuned. (e) Hinge modes (f) surface phase transition at $\mu^{c,1}$ (g) Majorana zero modes (h) vortex phase transition at $\mu^{c,2}$. $M=-2.5,t_1=1,t_2=1,\delta_0=0.3$. Lattices of size of $40\times 30\times 30$ are used to obtain the 3D probability density profiles.}
    \label{fig:SPT}
\end{figure*}
\emph{Results}.-- We obtain the low-energy spectrum in the presence of a vortex by numerically diagonalizing the BdG Hamiltonian in Eq.~(\ref{hambdg}) with periodic boundary conditions in the direction parallel to the vortex line (results are shown in Fig.~\ref{fig:VPTvsk}).
For easy comparison with earlier results \cite{hosur2011} we choose the following set of parameters: $M=-2.5,\,t_0=t_1=1$
(in the remainder all energies are in units of $t_0$).
For this choice of parameters all of the interesting physics is happening near the $\Gamma$ point so we focus on $k_z=0$ ($k_x=0$) for vortex lines parallel to $z$ ($x$).

 For the parameters mentioned above, Ref.~\onlinecite{hosur2011} showed MZMs at the ends of vortex lines survive up to a critical doping of  $\mu^c\sim 0.9$ where the vortex lines undergo a VPT. In Fig.~\ref{fig:VPTvsk}(a), we have reproduced this result for comparison. For chemical potentials smaller than $\mu^c$ the spectrum on the (periodic) vortex line is gapped, and MZMs will appear if the vortex line is terminated at a surface. From symmetry this result is independent of the orientation of the vortex line, e.g., vortex lines parallel to $x$ or $z$ have the same $\mu^c.$

Next, we turn on $t_2$ to push the system into the HOTI phase and obtain the spectrum for vortex lines oriented in the $z$ (Fig.~\ref{fig:VPTvsk}(b)) and the $x$ (Fig.~\ref{fig:VPTvsk}(c)) directions. We find that a critical $\mu$ still exists around $\mu\sim 1.0$, for both vortex orientations, having only a small increase compared to the TI case. However, we see a qualitative difference for small chemical potentials, i.e.,  there are other gapless points in Fig.~\ref{fig:VPTvsk}(b,c) that are absent in the case of an ordinary TI. Without a vortex, a HOTI with superconducting pairing as in Eq.~(1), is gapped on all surfaces. However, it possesses low-energy Majorana hinge states for a specific range of $\mu$. In Fig. \ref{fig:VPTvsk}(b) the gapless point at $k_z=0$ and $\mu=0$ is the hinge modes of the proximitized HOTI. As $\mu$ is tuned away from zero the hinge states parallel to $z$ have their zero-energy nodal point shifted off $k_z=0$ so we only see finite-energy states associated to the hinges for $\mu\neq 0$. These hinge modes exist up to around $\mu\sim 0.3$ where there is a surface phase transition (See Fig.~S1 in SM).

Lets now consider a vortex line parallel to $x$ the corresponding spectrum for $k_x=0$ is shown in Fig.~\ref{fig:VPTvsk}(c). Up to $\mu^{c,1}$ (denoted by green vertical dashed line), there are hinge modes, and for this orientation the zero-energy nodal point is not shifted away from $k_x=0$ as $\mu$ is increased. At $\mu^{c,1}$ there is a surface phase transition, after which the hinge states are removed. For $\mu^{c,1}<\mu<\mu^{c,2}$ the 3D superconducting HOTI is not topological and the only zero energy states are
the Majoranas at the ends of the vortex. We show the evolution of the low-energy surface states (without a vortex) in  Fig.~\ref{fig:SPT}(a-c) as we tune $\mu$ through $\mu^{c,1}.$ In Fig. \ref{fig:SPT}(e-h) we show the 3D probability densities of low-energy modes when a vortex is present and is oriented along the $x$ direction. We first see hinge-states (e) followed by states extended along the surface (f) at the surface phase transition. After the surface phase transition the MZMs appear (g) which are subsequently destroyed at $\mu^{c,2}$, (h). Remarkably, we see that the MZMs appear only after the surface phase transition. This is in sharp contrast to the surfaces of the TI and the (001) surface of the HOTI as these always host MZMs as long as $\mu<\mu^{c,2}$.

To get a clearer understanding of the surface phase transition demonstrated in Fig.~\ref{fig:SPT}, we derive an effective surface Hamiltonian for the (100) surface of the superconducting HOTI by treating the $C^z_4\mathcal{T}$ symmetric term in Eq.~(\ref{eqhoti}) perturbatively. We find,$h_{eff}(\vex{k})=v_z k_z \tau_z\sigma_y+v_y k_y \tau_0\sigma_z-\frac{t_2}{2}\tau_z\sigma_x+\delta_0\tau_y\sigma_y-\mu\tau_z\sigma_0.$ We see that both the $C_4\mathcal{T}$ symmetric term, and the superconducting pairing, can independently gap out the surface states (since they both fully anticommute with the kinetic energy terms), but they commute with each other, therefore they can be thought of as \emph{competing} masses on the surface.
This can lead to a gap closure at $k_y=k_z=0$ when $\mu^c=\sqrt{t_2^2-4\delta^2_0}/2.$
%
% CHANGE
%
This is a general condition, therefore we expect that a surface transition will occur in any other model of chiral HOTIs
for which the surface projection of the mass and superconducting pairing term commute.
Similarly, for 3D HOTIs constructed by gluing together lower-dimensional topological phases \cite{Songeaax2007,khalafPRX2018,Khalaf2051} to
have a surface phase transition the mass terms of the 2D surfaces connected by the vortex must commute
with the superconducting pairing term.
A prediction from this analysis is that a surface phase transition only appears in the $t_2>2\delta_0$ regime, i.e., when the local $\mathcal{T}$-breaking surface gap is stronger than the proximity-induced gap. To show this more rigorously we compute $\mu^{c,1}$ numerically by varying $t_2$ for a fixed $\delta_0$. From Fig.~\ref{fig:SPT}(d), we clearly see that, for a fixed superconducting gap $\delta_0$, when increasing $t_2,$ $\mu^{c,1}$ increases in qualitative agreement with the analytical result.

We now show that, in addition to the chemical potential, an external Zeeman term can provide some amount of tunability of the surface phase transition. Let us consider a vortex line along the $x$ direction passing through the center of the $y-z$ plane. An external magnetic field $B_x$ in the $x$ direction gives rise to the Zeeman term
$B_x J_x=B_{x}\tau_z\kappa_0\sigma_x.$
This term can partially suppress the effect of $t_2$. $B_x=0.1$, for example, reduces $\mu^{c,1}$ from $~0.33$ to $~0.16$. For sufficiently strong magnetic field, the phase appears after the surface phase transition can still possess Majorana hinge states\cite{elsewhere}. A Zeeman field perpendicular to the vortex line (i.e., parallel to the surface) will not  influence $\mu^{c,1}$. Therefore, we find that  we can tune the surface phase transition on a given surface by applying a Zeeman field. If one has control over the orientation of the field one may be able to selectively tune the critical doping levels on each surface.

\begin{figure}[t]
    \centering
    \includegraphics[width=0.5\textwidth]{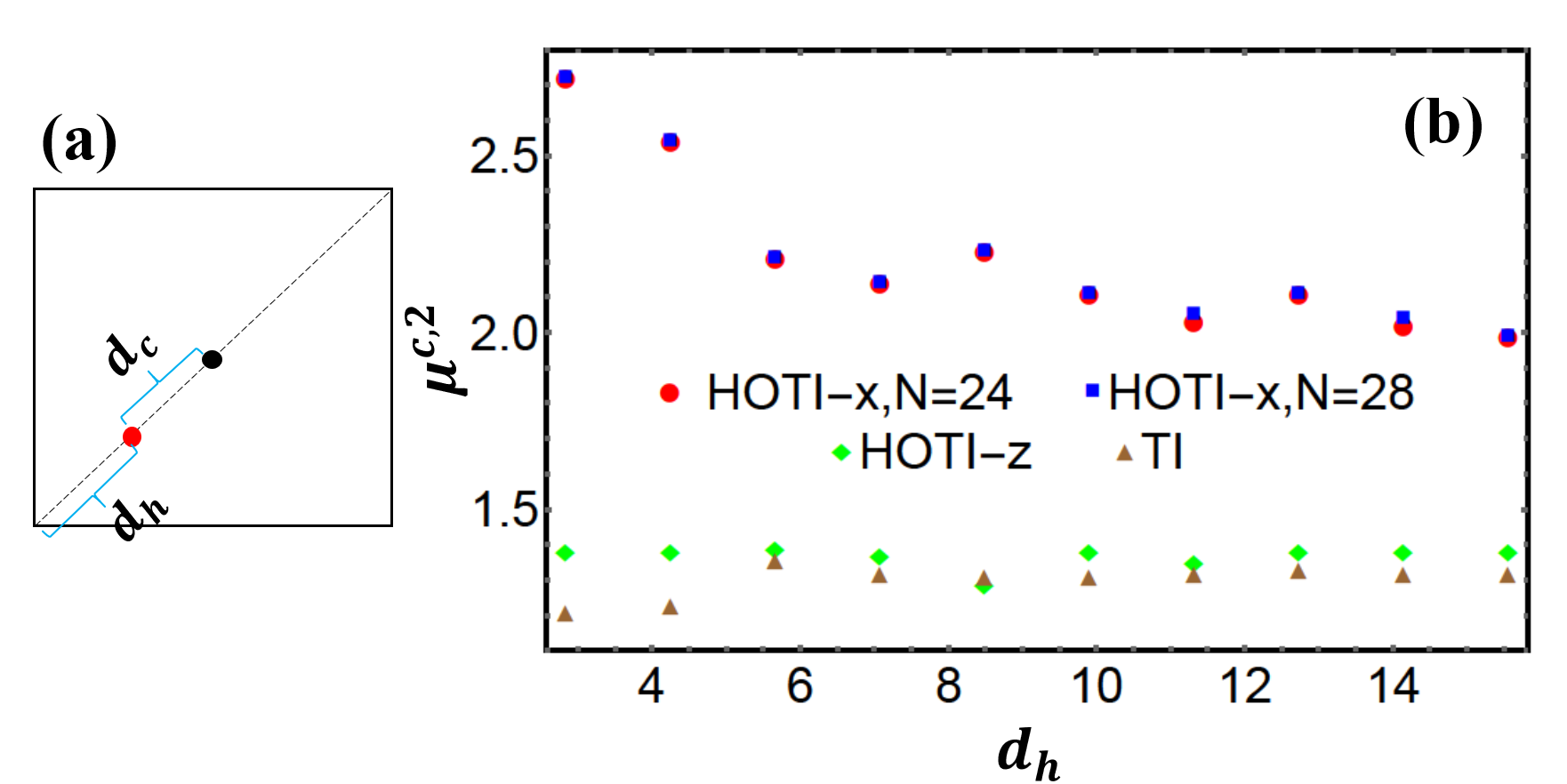}
    \caption{(a) schematic picture shows the distance between a vortex (red dot) from hinge $d_h$ and center $d_c$ of lattice (black dot), (b)$\mu^{c,2}$ vs $d_h$ for HOTI with superconducting vortex along $x$-direction (HOTI-$x$) with two system sizes of $N=24,28$ and HOTI-$z$ and TI with $N=28$. $M=-2,t_1=1,t_2=1 (\text{HOTI}),\delta_0=0.2$. }
    \label{fig:Mucvsdh}
\end{figure}

Now, we briefly discuss how the location of the vortex center ($V_c$) on the side surfaces affects the critical doping $\mu^{c,2}$. Let the distance between the vortex center and the hinge at corner $(1,1)$ be $d_h=(n-1)\sqrt{2},$ and the distance between the vortex center and the center of the plane be $d_c=(N/2-n)\sqrt{2}$, where the surface is an $N\times N$ square, and $n$ is the coordinate of the vortex center along the diagonal (Fig.~\ref{fig:Mucvsdh}(a)). Fig.~\ref{fig:Mucvsdh}(b) shows $\mu^{c,2}$ versus $d_h$, for two different system sizes, $N=24, 28$. By changing, the system size, $d_c$ changes but $d_h$ remains same. By decreasing $d_h$ we find (unlike the TIs or HOTIs with a vortex parallel to the $z$ direction) $\mu^{c,2}$ increases and the results are independent of $N$, therefore we conclude that this is not a finite-size effect and that the vortex is sensitive to $d_h$. The details of the variation (and how strong the variation is), are a function of various parameters of the system, and in some cases can be affected more strongly by finite size effects. We note that for a fixed set of system parameters, the variation of $d_h$ only affects $\mu^{c,2}$ and not $\mu^{c,1}$ as the surface physics is nominally insensitive to the presence of the vortex. However, an external Zeeman term could affect the results of Fig.~\ref{fig:Mucvsdh}(b) indirectly, since the presence/absence/spatial profile of the hinge states are effected by the Zeeman field, and thus the vortex-hinge hybridization can be affected by such a field. This effect may be difficult to observe with a single (likely pinned) vortex, but may be measureable with a vortex lattice where one might expect a distribution of $\mu^{c,2}$ across the lattice depending on the spatial location of the vortices relative to the hinges.

In summary, we have found that vortex phenomenology can be utilized to distinguish a class of HOTIs from TIs, and can also serve as a platform for the realization of Majorana zero modes.
In HOTIs having proximity-induced or intrinsic s-wave superconductivity we identified a new critical doping $\mu^{c,1}$ that marks a topological surface phase transition for the gapped surfaces. We showed that MZMs only appear in a range of doping levels between $\mu^{c,1}$ and $\mu^{c,2},$ the latter being the critical doping for the known vortex phase transition.
The surface transition results from two competing mass terms on the gapped surfaces of HOTI: the superconducting gap and a $\mathcal{T}$-breaking mass term resulting from the bulk $C_4\mathcal{T}$ symmetric term that is responsible for driving a parent $\mathcal{T}$-invariant TI into the HOTI.
%
% CHANGE
%
We note that the latter condition is general and surface phase transition should occur in any chiral HOTIs
for which the projection on the surfaces connected by the vortex of the mass term and of the superconducting pairing term commute.
In the SM we consider two other HOTI models to exemplify this general condition.

The surface phase transition can be tuned with chemical potential or Zeeman fields which may help the experimental detection of the phase transitions and MZMs.
Recently, iron-based superconductors attracted attention because they can host both topological and trivial vortices simultaneously \cite{vortexexp1,vortexexp2,vortexexp3,vortexIron1,vortexIron2,vortexIron3,Iron1,vortexIron4,Ghazaryan2019,Iron2}.
Here we have shown that the HOTIs are another example of such systems, which for a specific range of dopings and applied Zeeman fields, can host both trivial and topological vortices, but in different orientations.
We have provided numerical evidence showing strong variation of the critical doping $\mu^{c,2}$ depending on the spatial location of the vortex center.
%
% CHANGE
%
Very recently several materials, and heterostructures, have been proposed as possible chiral HOTIs:
EuIn$_2$As$_2$ \cite{exhoti1}, EuSn$_2$As$_2$ \cite{exhoti7}, MnBi2Te$_4$ \cite{exhoti3,exhoti4,exhoti5,exhoti6,exhoti7}, and CrI$_3$/Bi$_2$Se$_3$/MnBi$_2$Se$_4$ heterostructures \cite{exhoti2}.
A possible setup to observe our theoretical predictions could be realized
by combining one of the materials above in a heterostructure consisting of an s-wave superconductor
and and an external gate (separated by the HOTI by a high quality dielectric) to tune the chemical potential of the HOTI,
in which a vortex is induced via an external magnetic field along the $x$ direction (see also SM).

Acknowledgments.-- S. A. A. G. and E. R. acknowledge support from ARO
(Grant No. W911NF-18-1-0290), NSF (Grant No. DMR1455233). E. R. also thanks the Aspen Center for Physics,
which is supported by NSF Grant No. PHY-1607611,
where part of this work was performed. The authors
acknowledge William $\&$ Mary Research Computing for
providing computational resources that have contributed to
the results reported within this Letter. T. L. H. thanks the
U.S. National Science Foundation under Grant No. DMR
1351895-CAR, and the MRSEC program under NSF Grant
No. DMR-1720633 (SuperSEED) for support. T. L. H. and
E. R. also thank the National Science Foundation under
Grant No. NSF PHY1748958(KITP) for partial support at
the end stage of this work during the Topological Quantum
Matter program and Spin and Heat Transport in Quantum
and Topological Materials, respectively.

%\bibliographystyle{revtex4}
%\bibliography{VPT_HOTI_03_shortened}


%%%%%%%%%% Merge with supplemental materials %%%%%%%%%%
\pagebreak
\widetext
\begin{center}
\textbf{\large Supplementary Materials: Surface and vortex phase transition in superconducting higher-order topological insulators}
\end{center}
%%%%%%%%%% Merge with supplemental materials %%%%%%%%%%
%%%%%%%%%% Prefix a "S" to all equations, figures, tables and reset the counter %%%%%%%%%%
\setcounter{equation}{0}
\setcounter{figure}{0}
\setcounter{table}{0}
\setcounter{page}{1}
\makeatletter
\renewcommand{\theequation}{S\arabic{equation}}
\renewcommand{\thefigure}{S\arabic{figure}}
\renewcommand{\bibnumfmt}[1]{[S#1]}
\renewcommand{\citenumfont}[1]{S#1}
%%%%%%%%%% Prefix a "S" to all equations, figures, tables and reset the counter %%%%%%%%%%

%%%%%%%%%%%%%%%%%%%%%%%%%%%%%%%%%%%%%%%%%%%%%%%%%%%%%%%%%%%%%%%
\section{Hinge spectrum of Superconducting HOTIs}
The hinge spectrum of superconducting HOTI given by Eqs.~(1) and (2) of the main text along the $x$ direction corresponding to zero modes of Fig.2c.

%\note{Change all the figure numbers to S1, S2,...}.

\begin{figure}[H]
\centering
\includegraphics[width=0.7\textwidth]{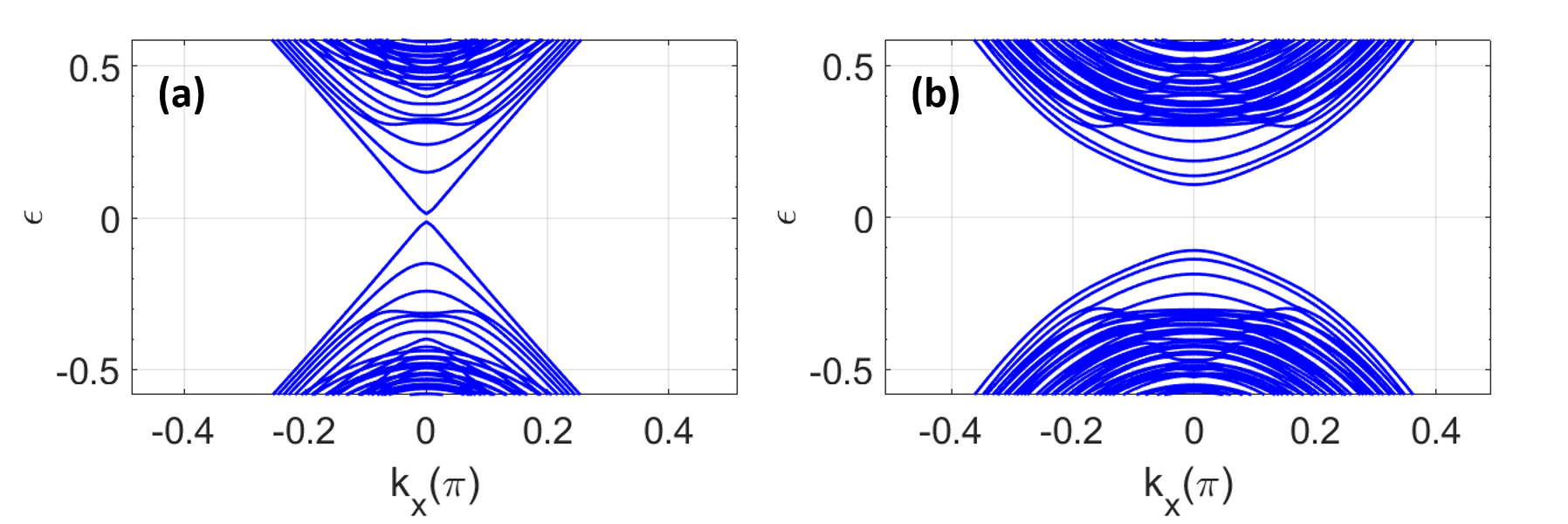}
\caption{The hinge spectrum along the $k_x$ direction (a) before [$\mu=0.2$] and (b) after [$\mu=0.5$] the surface phase transition. The tiny gap in (a) is finite-size effect. The systems size of $100\times100$ is used.}
\label{Fig:hingekx}
\end{figure}

\section{Two more models of chiral higher-order TIs }

Here we present the VPT spectrum for two more models of chiral HOTIs \cite{choti1,choti2,choti3,choti4,choti5,choti6,choti7}. The first is protected by a combination of fourfold rotation and inversion symmetries, i.e $C_4\mathcal{I}$. This model differs from the model in the main text by a simple perturbation, namely an extra magnetic field, $h$, along the $(0,0,1)$ direction:
\begin{align}\label{c4Isymmetric}
h_{c_4\mathbb{I}}(\vex{k})=\,\left(M+t_0\sum_i \cos (k_i)\right)\kappa_z\sigma_0+\,t_1\sum_i\sin (k_i)\kappa_x\sigma_i\,+\,t_2\large(\cos (k_x) - \cos (k_y)\large)\kappa_y\sigma_0\,+\,h\kappa_0\sigma_z.
\end{align}

On the other hand one can build a $\mathcal{I}$ symmetric HOTI by turning off the $C_4\mathcal{T}$ symmetric term ,$t_2=0$, but instead applying the magnetic field in all three directions to gap all the surfaces as,

\begin{align}\label{Isymmetric}
h_{\mathbb{I}}(\vex{k})=\,\left(M+t_0\sum_i \cos (k_i)\right)\kappa_z\sigma_0+\,t_1\sum_i\sin (k_i)\kappa_x\sigma_i\,+\,\kappa_0\bold{h}\,.\,\mathbf{\sigma}.
\end{align}

\begin{figure}[H]
\centering
\includegraphics[width=0.7\textwidth]{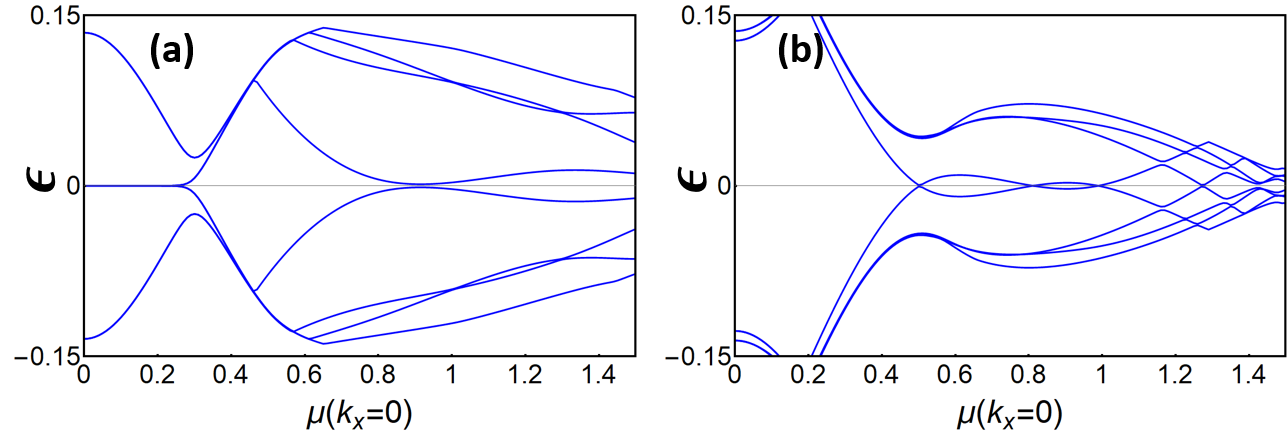}
\caption{The VPT spectrum of a HOTI at $kx = 0$ having a vortex line parallel to $x$ for models of (a) Eq.~(\ref{c4Isymmetric}), (b) Eq.~(\ref{Isymmetric}). $M=-2.5,\,t_1=1,\,t_2=1,\,\delta_0=0.3$ and $h=0.15$ is used for all plots.}
\label{Fig:VPTC4IandI}
\end{figure}

Fig~.\ref{Fig:VPTC4IandI} shows the VPT spectrum for model of Eq.~\ref{c4Isymmetric} (Fig.~\ref{Fig:VPTC4IandI}a) and Eq.~\ref{Isymmetric} (Fig.~\ref{Fig:VPTC4IandI}b), respectively. As we discussed in the main text a magnetic field parallel to the surface which vortex is passing through, does not affect the vortex phenomena. Therefore as we expected the physics discussed in the text remains valid for the $C_4\mathbb{I}$ symmetric model of Eq.~(\ref{c4Isymmetric}) along the $x$-direction. Of course, the magnetic field along the $z$-direction will gap out the $z$-surface, however, it does not lead to a surface phase transition on that surface. The Fig.~\ref{Fig:VPTC4IandI}b, shows the VPT spectrum along the $x$-direction for the case of the model of Eq.~(\ref{Isymmetric}). We observe that in the absence of $t_2$ term, the magnetic field alone would not lead to a surface phase transition (at least for the case with simple spin-singlet s-wave pairing) and MZMs exist up to vortex phase transition point.
% We leave further investigation all possible pairings and their possible topological phases for future works.

\section{Other Superconducting pairings}

In the main text, we focused on the spin-singlet intra-orbital pairing which has the most experimental relevance. Here we show the VPT spectrum for five other possible $k$-independent pairings allowed by Fermi statistics. However, we do not intend to analyze the physics in depth as each
pairing deserves
to be discussed separately elsewhere.
Instead, we show the the spectrum and point out some qualitative remarks. We can write the BdG Hamiltonian as,

\begin{eqnarray}\label{hambdg}
h^{BdG}_{\vex{k}}=\begin{bmatrix}
   H_{HOTI}(\vex{k})-\mu & \delta_0\Delta_i \\
   \delta_0\Delta^*_i & -H^T_{HOTI}(-\vex{k})+\mu\\
\end{bmatrix}
\end{eqnarray}

The requirement that the pairing term satisfies Fermi-Dirac statistics implies that there are only six possible pairing matrices $\Delta_i$ (See e.g., \cite{Roy-Ghorashi,Fu-Berg,Sato}): (1) $\Delta_1=\kappa_0\sigma_y$, (2) $\Delta_2=\kappa_z\sigma_y$, (3) $\Delta_3=\kappa_x\sigma_y$, (4) $\Delta_4=\kappa_y\sigma_x$, (5) $\Delta_5=\kappa_y\sigma_z$, (6) $\Delta_6=\kappa_y\sigma_0$. In the main text we examined the simplest case of intra-orbital spin-singlet pairing of $\Delta_1$. This is exactly the same paring which is used in the original work of \cite{pavan-Ashvin}.

\begin{figure}[H]
\centering
\includegraphics[width=0.7\textwidth]{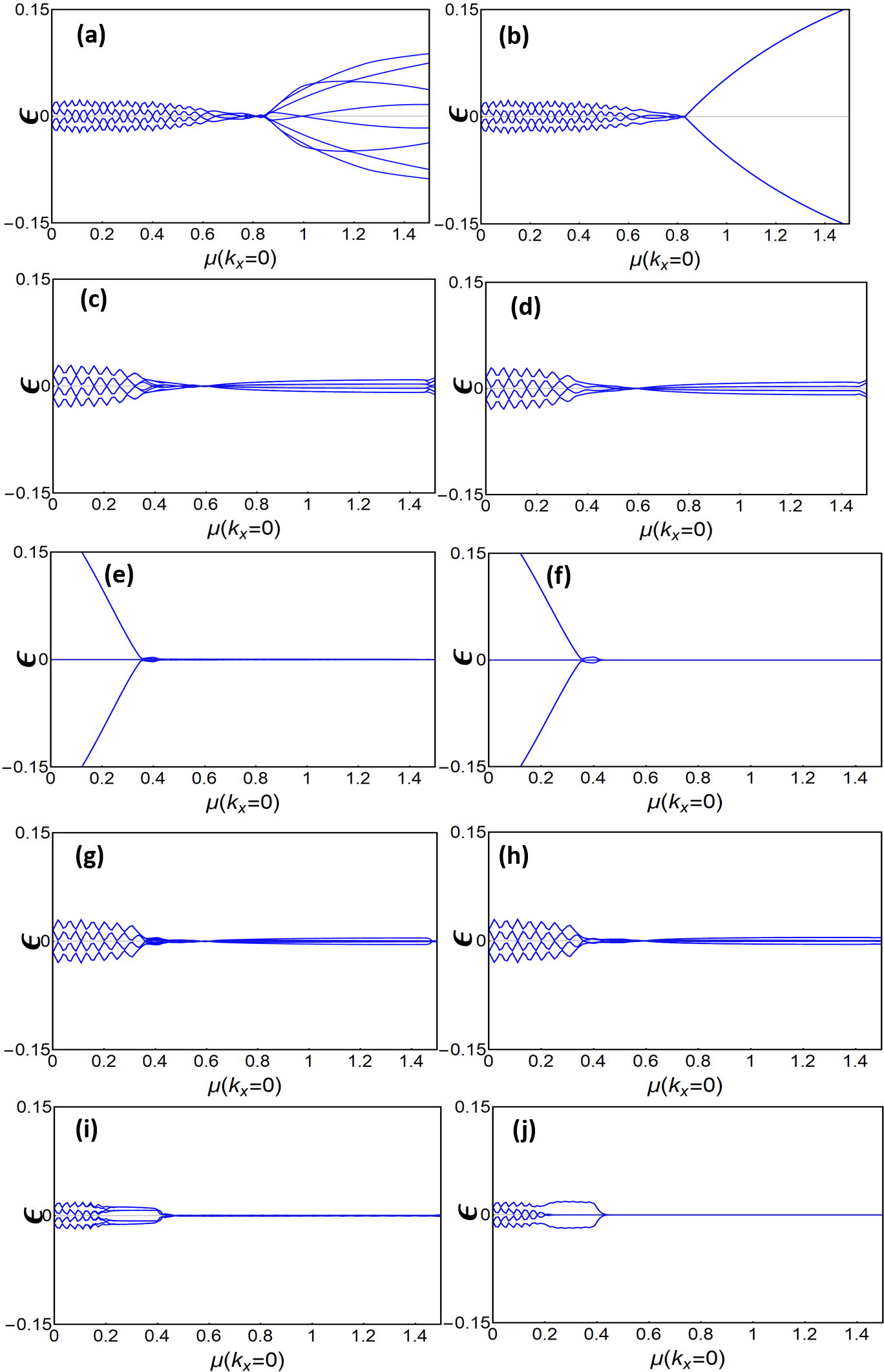}
\caption{The VPT spectrum of a HOTI at $kx = 0$ having a vortex line parallel to $x$ for (a) $\Delta_2$, (c) $\Delta_3$, (e) $\Delta_4$, (g) $\Delta_5$, (i) $\Delta_6$. The second column (b,d,f,h,j) shows the spectrum of corresponding pairings in the absence of vortex.$M=-2.5,\,t_1=1,\,t_2=1,\,\delta_0=0.3$ is used for all plots.}
\label{Fig:VPTallpairing}
\end{figure}

The results shown Fig.~\ref{Fig:VPTallpairing} were obtained using the same parameters, and system sizes, as the ones shown in Fig.2c of the main text.
To facilitate the comparison we also plotted them in the same energy range. Now, we make few qualitative remarks. First, comparing to the model $\Delta_1$ of the main text, all the plots of Fig.~\ref{Fig:VPTallpairing}, show many gapless points and often in a very narrow window of energies.
This fact makes
the analysis of the spectra and the identification of the nature of the low (zero) energy modes extremely difficult.
A full understanding of the physics for these pairings would require extensive full three dimensional
calculations, that we feel are beyond the scope of the current work.
%
%This complicates the full analysis and of the distinguishing the surface, hinges and vortex modes. Moreover, in the case of appearance of MZMs, their %detection can be obscured due to other modes in the system. Second, in the absence of a vortex, there are distinct features for the pairings $\Delta_i\neq\Delta_1$ which qualitatively distinguish them from the $\Delta_1$ discussed in the main text. For example,
For $\Delta_4,\,\Delta_5,\,\Delta_6$, the bulk spectrum become gapless at some finite doping $\mu$ which results in Majorana Fermi arcs on some of the surfaces. For $\Delta_2$ and $\Delta_3$ the system undergoes a bulk phase transition (bulk gap closure) in sharp contrast to the $\Delta_1$ of the main text. Because these  we can have  zero modes in the bulk, surface, and possibly hinges of the system, modes that and would affect the vortex physics.

Finally, we note, despite the all of the differences and difficulties mentioned here, it does not mean that they are less interesting (at least theoretically), but they deserve separate thorough investigation.

\section{Possible Experimental Setup}
\begin{figure}
\centering
\includegraphics[width=0.5\textwidth]{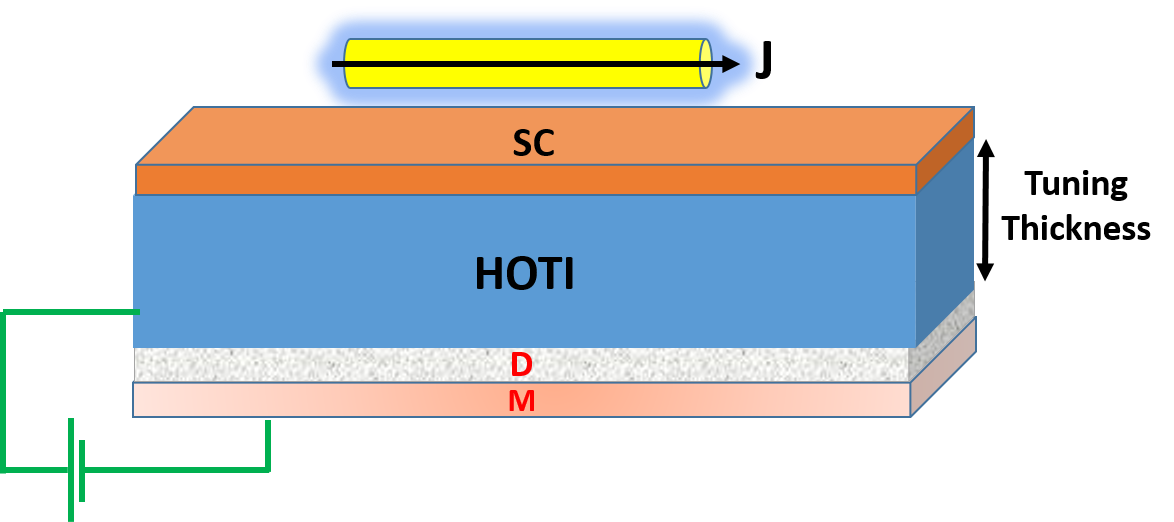}
\caption{A schematic sketch of a possible experimental setup. The "D" and "M" stands for dielectric and metal, respectively.}
\label{expschematic}
\end{figure}

Our findings could be directly relevant for potential chiral HOTIs and axion insulators candidates, such as in EuIn2As2 \cite{S_exhoti1}, CrI3/Bi2Se3/MnBi2Se4 heterostructures \cite{S_exhoti2}, MnBi2Te4 \cite{S_exhoti3,S_exhoti4,S_exhoti5,S_exhoti6,S_exhoti7} and EuSn2As2 \cite{S_exhoti7}, all of which are predicted to host chiral hinge states.
A possible setup to observe our theoretical predictions could be realized
by combining one of the materials above in a heterostructure consisting of an s-wave superconductor
and and an external gate (separated by the HOTI by a high quality dielectric) to tune the chemical potential of the HOTI,
in which a vortex is induced via an external magnetic field along the $x$ direction,
as shown schematically in Fig.~\ref{expschematic}.
%
%A heterostructure in which one of the systems above is proximitized to a superconductor could be ideal to realize the physics discussed in our work. %Fig.~\ref{expschematic}, shows a schematic sketch of a possible setup to realize a superconducting HOTI and how to induce a vortex in it. In this setup an external gate (separated by the HOTI by a high quality dielectric) could be used to tune the chemical potential of the HOTI, and the vortex is induced via an external magnetic field along the x direction.

\end{document}